\documentstyle[12pt,aasms4]{article}

\begin{document}

\title{LARGE-SCALE REGULAR MORPHOLOGICAL PATTERNS
       IN THE RADIO JET OF NGC 6251}

\author{Hiroshi Sudou \& Yoshiaki Taniguchi}

\affil{Astronomical Institute, Graduate School of Science,
       Tohoku University, Aoba, Sendai 980-8578, Japan}

\begin{abstract}
We report on large-scale,
regular morphological patterns found in the radio jet of the
nearby radio galaxy NGC 6251.
Investigating morphological properties of this radio jet
{}from the nucleus to a radial distance of $\sim$ 300 arcsec 
($\approx$ 140 kpc) mapped at 1662 MHz and 4885 MHz 
by Perley, Bridle, \& Willis, we find three chains, each of which 
consists of five radio knots. We also find that eight radio knots in the
first two chains consist of three small sub-knots (the triple-knotty
substructures). We discuss the observational properties of 
these regular morphological patterns.
\end{abstract}


\section{INTRODUCTION}

Since the discovery of the powerful radio jet from quasar 3C273
(Hazard et al. 1963, see for a review Bridle \& Perley 1984;
Zensus 1997), 
the generation of radio jets has been
one of the long-standing main problems in active galactic nuclei
(e.g., Begelman, Blandford, \& Rees 1984; Urry \& Padovani 1995). 
The most probable energy sources have been considered
to be either mass-accreting, supermassive, single black holes around which
gravitational energy is transformed into huge kinetic and radiation
energies with the help of gaseous accretion disks (e.g., Rees 1984)
or the electromagnetic extraction of energy from spinning supermassive 
black holes (Blandford \& Znajek 1977; see also Wilson \& Colbert 1995).
In order to understand the genesis of radio jets, it is important
to find some reliable observational constraints on theoretical models.

It has been noticed that parsec-scale radio jets probed by
VLBI techniques often show wiggles (e.g., Whitmore \& Matese 1981; Roos 1988;
Roos, Kaastra, \& Hummel 1993). 
If a certain mechanism responsible for the formation of the wiggles
(e.g., the precession of a supermassive black hole binary;
Begelman, Blandford, \& Rees 1980)
has been working since the onset of radio jet activity, there could be some
morphological evidence even in well-developed (i.e., $\sim$ 100 kpc scale)
radio jets. Furthermore, 
recent progress in three-dimensional numerical MHD simulations
has enabled us to examine the detail of morphological properties
of radio jets due to the so-called Kelvin-Holmheltz instability 
in the magnetic fluid (e.g., Rosen et al. 1998; Koide, Shibata, \&
Kudoh 1999).  Therefore,
since morphological properties of actual radio jets
provide important constraints on the physical process involved in 
radio jets, it is interesting
to investigate overall morphological properties of some well-developed
radio jets in detail.

For this purpose, in this paper, we investigate morphological
properties of the radio jet of NGC 6251 in detail because
the radio jet of NGC 6251 is one of the brightest known examples of
a well-developed jet (Waggett, Warner, \& Baldwin 1977;
Cohen \& Readhead 1979; Perley, Bridle, \&
Willis 1984; Jones et al. 1986; Jones \& Wehrle 1994).
We use a distance to NGC 6251, 94.4 Mpc,
which is determined with the use of a recession velocity of NGC 6251
to the galactic standard of rest, $V_{\rm GSR}$ = 7079 km s$^{-1}$
(de Vaucouleurs et al. 1991), and a Hubble constant,
$H_{0}$ = 75 km s$^{-1}$ Mpc$^{-1}$.

\section{RESULTS}

\subsection{A Large-scale Regular Morphological Pattern}

First, we investigate large-scale morphological properties of 
the radio jet of NGC 6251 using the radio continuum image at 
4885 MHz obtained by Perley et al. (1984; see the right panel
of Figure 5 in their paper). In the top panel of  Figure 1, we show
the image from the radio nucleus to a radius of $\sim$ 300 arcsec
($\approx$ 140 kpc). 
The spatial resolution of this image is 4.4 arcsec. 
There can be seen more than a dozen knots along the radio jet.
We find that there are three sets of radio knots; hereafter
Chain A, B, and C (see Figure 1).
Each chain appears to consist of five knots although the first knot
in Chain A can be marginally discerned because it is so close to 
the radio nucleus. The fifth knot in each Chain is slightly displaced
outward with respect to the inner four knots.

In order to examine whether or not these Chains show some regular
patterns, we measure the positions of 15 knots; the measurements 
are made using a scale on an enlarged photocopy of the image
shown in Figure 1. The measurement error is $\approx$ 0.5 mm 
on the photocopy, corresponding to $\approx$ 0.6 arcsec; note that 
the actual measurement error is about 0.1 mm on the photocopy
but the ambiguity of the peak of radio knots is estimated to be
$\approx$ 0.5 mm.
The following quantities are measured in our analysis
(see Figure 2).
1) $d$ and $D$: the radial distance
in units of arcsec and kpc respectively, 2) $s$ and $S$: 
the separation between
adjacent knots in units of arcsec and kpc respectively, and
3) $w$ and $W$:
the separation of the $i$-th knot between the adjacent Chains
in units of arcsec and kpc respectively. 
The results are summarized in Table 1.

In Figure 3, we show the angular distances ($d$) of the fifteen knots
found in three Chains A, B, and C. It is shown that the knots do 
not develop linearly in distance; the five knots in Chain C
are more distant than the position extrapolated using the positions
of the five knots in both Chains A and B.
In order to see this property quantitatively, we show
the separations of the $i$-th knot between two different Chains
($w$) in Figure 4. The separations between Chains B and C [$w$(B-C)]
are systematically larger than those between Chains A and B
[$w$(A-B)]; i.e., $w$(A-B) $\simeq$ 67 -- 89 arcsec while $w$(B-C)
$\simeq$ 120 -- 132 arcsec, corresponding to linear
separations of $W$(A-B) $\simeq$ 42 -- 55 kpc
and $W$(B-C) $\simeq$ 75 -- 82 kpc.

In Figure 5, we show the angular separation between two adjacent knots.
The separation appears to be increasing with increasing distance
although there can be seen some scatters in the data points.

Another interesting property found in Figure 1 is that the separations
among the five knots appear to be larger for the outer Chain.
In Figure 6, we show the separations between two adjacent knots ($s$)
in three Chains separately. The separations increase from Chain A
through B to C. It is also found  that the variation 
patterns in $s$[($i$+1) - $i$] from $i$ = 1 to 4 are quite similar 
among three Chains. These findings reinforce that our identifications
of three Chains are really meaningful.
It is thus suggested that the radio jet activity in NGC 6251 is not
sporadic but regular with some periodicity.

\subsection{Substructures in Radio Knots}

We further investigate detailed morphological properties of the
individual radio knots. To perform this, we use the radio continuum
image at 1662 MHz given by Perley et al. (1984; see Figure 4
in their paper).
This map covers the inner 2 arcmin ($\approx$ 55 kpc)
region of the radio jet
with the spatial resolution of 1.15 arcsec.
In Figure 7, we show the map together with close-up images of eight
individual knots; the first four knots are in Chain A 
(A2, A3, A4, and A5) and the remaining four knots are in Chain B
(B1, B2, B3, and B4). These identifications are the same as 
those in section 2.1 (see Table 1).
Each knot appears to consist of three brightness peaks; 
hereafter the triple-knotty
substructure. In particular, this substructure is unambiguously seen in 
the first two knots, A2 and A3.

\section{DISCUSSION}

We have shown that there are two kinds of regular morphological
structures in the radio jet of NGC 6251; 1) three Chains 
consisting of five radio knots , and 2) the triple-knotty
substructures in the individual knots.
Here we discuss some possible origins of these large-scale
regular structures. 

\subsection{Chains}

The presence of three Chains suggests that a certain periodicity
is involved in the radio jet activity of NGC 6251.
As shown in Table 1, the separations of the $i$-th knots ($W$)
are quite similar between two adjacent Chains; i.e., the average values
are $\overline{W}$(A-B) $\simeq 48.0 \pm 0.5$ kpc and
$\overline{W}$(B-C) $\simeq 79.3 \pm 2.9$ kpc.
We estimate timescales corresponding to these separations.
In order to perform this, both the viewing angle toward the jet
($\theta_{\rm jet}$) and the jet velocity ($v_{\rm jet}$) are
necessary for the kpc-scale radio jet.

Since the parsec-scale counterjet cannot be seen in the previous
VLBI observations (Perley et al. 1984; Jones et al. 1986; Jones
\& Wehrle 1994; see however Sudou et al. 2000a, b),
it is unlikely that the radio jet of NGC 6251 lies close to
the celestial plane. Jones et al. (1986) estimate that
the angle between the radio jet and our line of sight may be
$\theta_{\rm jet} \sim 45^\circ$ based on the observed 
jet-to-counterjet intensity ratio.  We therefore adopt
$\theta_{\rm jet} = 45^\circ$ in later analysis.
Another important quantity is the large-scale
(i.e., kpc-scale) jet velocity $v_{\rm jet}$
which is also difficult to be estimated (e.g., Perley et al. 1984).
Based on several constraints (e.g., the energy flux required to 
power the radio jet, etc.), Perley et al. (1984) suggest that 
the large-scale jet velocity of NGC 6251 is subrelativistic; $v_{\rm
jet} \leq 0.1c$. 
It is known that ram pressure confinement for
the strongest double-lobed radio sources such as NGC 6251 requires
$v_{\rm jet} \simeq 0.1c$ (e.g., Begelman et al. 1984).
On the other hand,
using the observed jet-to-counterjet brightness ratio, 
Jones et al. (1986) suggests  $v_{\rm jet} \cos \theta_{\rm jet} \geq
0.6$ for 
brightness ratio $ \geq 30$. Given $\theta_{\rm jet} = 45^\circ$,
they obtain $v_{\rm jet} \sim  0.84c$.
Since this estimate seems more reliable, we adopt for simplicity $v_{\rm
jet} = 0.8c$ 
in later analysis. 
These assumptions (i.e., $\theta_{\rm jet} = 45^\circ$ and
$v_{\rm jet} = 0.8c$) seem enough to estimate rough timescales
related to the large-scale regular structures found in this study.

First we estimate timescales related to three Chains A, B, and C.
Since the jet velocity is relativistic, we have to take account of
the relativistic aberration effect. The true projected distance
of the radio jet from the nucleus is estimated as 
$D_{\rm jet}' = \delta \ D_{\rm jet}$
where $\delta$ is the Doppler factor defined as
$\delta = [\gamma(1 - (v_{\rm jet}/c) \cos \theta_{\rm jet})]^{-1}$
where $\gamma = [1 - (v_{\rm jet}/c)^2]^{-1/2}$. 
Thus the true length of the radio jet is estimated as 
$L_{\rm jet}^0  = D_{\rm jet}' (\sin \theta_{\rm jet})^{-1}$. Therefore,
the related timescale $\tau_{\rm jet}$ is estimated as

\begin{eqnarray}
\tau_{\rm jet} & = & L_{\rm jet}^0 / v_{\rm jet} \nonumber \\
      & = & \ D_{\rm jet}'/
       (0.8 c ~ {\rm sin} \theta_{\rm jet}) \nonumber \\
      & \simeq & 1.83 \times 10^{11}\  \delta \ D_{\rm jet, 1}\ 
       v_{\rm jet, 0.8}^{-1}\ 
       (\sin \theta_{\rm jet, 45})^{-1}~ {\rm s} \nonumber \\  
      & \simeq & 5.79  \times 10^3\   \delta \ D_{\rm jet, 1}\ 
       v_{\rm jet, 0.8}^{-1}\ 
       (\sin \theta_{\rm jet, 45})^{-1}  ~ {\rm y}
\end{eqnarray}
where $D_{\rm jet, 1}$ is the jet length projected onto the
celestial plane in units of 1 kpc, $v_{\rm jet, 0.8}$ is the jet
velocity in units of $0.8c$, and $\theta_{\rm jet, 45}$ is the jet viewing
angle in units of 45$^\circ$. Given the
velocity and viewing angle assumed above, we obtain $\delta = 1.38$. The
projected jet lengths of three Chains are  
$D_{\rm jet}$(A) = $D_5$(A) $-$ $D_1$(A)
 = 32.4 $-$ 3.6 = 28.8 kpc,
$D_{\rm jet}$(B) = $D_5$(B) $-$ $D_1$(B)
 = 87.8 $-$ 45.5 = 42.3 kpc,
and
$D_{\rm jet}$(C) = $D_5$(C) $-$ $D_1$(C)
 = 169.9 $-$ 120.7 = 49.2 kpc.
Then we obtain the durations required to develop the radio jet for
three Chains; $\tau_{\rm jet}$(A) $\approx 2.3 \times 10^5$ years,
$\tau_{\rm jet}$(B) $\approx 3.4 \times 10^5$ years, and
$\tau_{\rm jet}$(C) $\approx 3.9 \times 10^5$ years.
These durations are shorten by one order of magnitude than 
the precession period estimated by Jones et al. (1986; see also
Begelman et al. 1980), 
$\tau_{\rm prec} \simeq 1.8 \times 10^6$ y. This precession is proposed
to explain the global wiggle pattern of the radio jet of NGC 6251.

As estimated above, the length of Chain becomes longer with increasing 
radial distance; i.e., $D_{\rm jet}$(A) $<$ $D_{\rm jet}$(B)
$<$ $D_{\rm jet}$(C). If this tendency is real, the radio jet of NGC 6251
must be accelerated even at several tens of kpc. An alternative idea may be 
that the radio jet is bending in a plane which encloses the radio 
jet and our line of sight to the jet. Since the position angle
of the parsec-scale radio jet (PA = 302$^\circ$.2 $\pm$ 0$^\circ$.8, for the
epoch of 1950.0;
Cohen \& Readhead 1979) is slightly different from that of the kpc-scale
jet (PA = 296$^\circ$.5; Waggett et al. 1977), Cohen \& Readhead (1979)
suggest a possible bending of the radio jet of NGC 6251.
Therefore, it is interesting to investigate this idea in more detail.

Here we assume that the true jet lengths of three Chains are nearly 
the same and the observed differences among them are attributed
to the differences in the viewing angle toward them.
If this is the case, we obtain the following relation;

\begin{equation}
{D_{\rm jet}({\rm A}) ~ \delta({\rm A})
 \over {\rm sin} \theta_{\rm jet}({\rm A})} =
{D_{\rm jet}({\rm B}) ~ \delta({\rm B})
 \over {\rm sin} \theta_{\rm jet}({\rm B})} =
{D_{\rm jet}({\rm C}) ~ \delta({\rm C})
 \over {\rm sin} \theta_{\rm jet}({\rm C})} 
\end{equation}
where $\theta_{\rm jet}({\rm A})$, $\theta_{\rm jet}({\rm B})$,
and $\theta_{\rm jet}({\rm C})$ are the average viewing angles
toward Chain A, B, and C, and $\delta(\rm A)$,  $\delta(\rm B)$, and 
$\delta(\rm C)$ are the average Doppler factors toward Chain A, B, and C,
respectively.
If we adopt $\theta_{\rm jet}({\rm C}) = 45.0^\circ$ and $\delta$(C) = 1.38,   
we obtain $\theta_{\rm jet}({\rm A}) \simeq 33.1^\circ$
and $\theta_{\rm jet}({\rm B}) \simeq 41.3^\circ$, with $\delta({\rm 
A}) \simeq 1.82$ and $\delta({\rm B}) \simeq 1.50$.  
It is therefore
suggested that the direction of the radio jet is approaching 
the line of sight as time goes by, or the jet flow follows fixed but
bent path. This result is schematically illustrated in Figure 8.

Recently, Sudou et al. (2000a, b) have found the counterjet at sub-parsec
scale and estimated the viewing angle to the sub-pc scale jet 
$\theta_{\rm jet} \simeq$ 17$^\circ$ -- 31$^\circ$. 
Since the viewing angle toward Chain A derived above is consistent
with their new estimate, this bending jet model appears
consistent with the observation. 
Therefore, it is not necessary to introduce the jet acceleration
at kpc regions.

\subsection{The Knots in Chains}

We investigate the separations of the knots in three Chains.
An average separation of the knots in each Chain is; 
$\overline{S} = \sum_{i=1}^4 S_{[i+1]-1}/4
\simeq 7.2 \pm 2.9$ kpc for Chain A, $10.6 \pm 4.5$ kpc for 
Chain B, and $12.3 \pm 3.1$ kpc for Chain C. Therefore, the 
average separation appears to increase with increasing Chain number.
It is noted that the fifth knot in each Chain is located at
a larger distance than that expected from the separations for
the remaining four knots (see column 5 of Table 1).
If we omit the data of the fifth knot, we obtain average separations
of $\overline{S} = \sum_{i=1}^3 S_{[i+1]-1}/3 \simeq 5.6 \pm 1.0$ 
kpc for Chain A,
$8.1 \pm 1.6$ kpc for Chain B, and $10.6 \pm 1.2$ kpc for Chain C.
Although the tendency  holds, the average separations are smaller
than the former estimates respectively.
If we adopt the bending jet model described in section 3.2 together 
with the relativistic aberration effect, the true separation,
$S^0 = \delta \overline{S}/ \sin \theta_{\rm jet}$, is; 
$S^0({\rm A}) \simeq 5.6 \times 1.82 / \sin 33.1^\circ 
\simeq 18.7 $ kpc, $S^0({\rm B}) \simeq 8.1 \times 1.50 /
\sin 41.3^\circ  
\simeq 18.4$ kpc, and  
$S^0({\rm C}) \simeq 10.6 \times 1.38 / \sin 45^\circ 
\simeq 20.7$ kpc. These values are similar to each other.

We obtain an average separation of the knots in three Chains
of $\simeq$ 19.2 kpc, corresponding to a timescale of
$\simeq 7.8 \times 10^4$ years given the jet velocity of
$v_{\rm jet} = 0.8c$.

\subsection{The Triple-Knotty Substructure}

We discuss briefly observational properties of the triple-knotty 
substructures. As shown in Figure 7, 
the first two knots (A2 and  A3) show a
clear triple-knotty substructures. However, the others show
a range of irregular, complex morphologies although we give possible 
identifications of the triple knots for them by arrows.
It is interesting to mention that such a triple-knotty substructure
is also found in the radio jet of Centaurus A (Clarke et al. 1986).

\subsection{Concluding Remarks}

In summary, the large-scale, regular morphological patterns 
involve the three kinds of structures; Chains, knots, and
triple-knotty substructures in the knots. The concerned
timescales for the first two structures are $\sim 10^5$ years
and $\sim 10^4$ years, 
respectively. Although the longest timescale obtained for
Chains may be related to the precession motion, the other two
timescales are not understood easily.

Although some studies of the large-scale morphological properties of the 
radio jets have been carried out (e.g., Perley \& Bridle 1984, Sparks et 
al. 1996, Perlman et al. 1999, Biretta et al. 1999, Bahcall et al. 1995, 
R$\ddot{\rm o}$ser et al. 1996, and Clarke et al. 1986), the analysis
presented in 
this paper is the first 
trial to investigate large-scale morphological regularity of radio
jets. Since this kind of analysis needs high-resolution and
large-scale radio continuum mapping, it seems difficult to perform
a systematic morphological study of radio jets. 
Therefore, at present, it is difficult to judge whether or not
large-scale morphological patterns found in the radio jet of NGC 6251
are general properties. However, this kind of analysis will be 
important  to provide observational constraints on the theory for
radio jets.

\vspace{0.5cm}

We thank to Dr. D. Jones and anonymous referee for their useful
comments and suggestions. 
This work was financially supported in part by Grant-in-Aids for the Scientific
Research (Nos. 10044052, and 10304013) of the Japanese Ministry of
Education, Culture, Sports, and Science.


\newpage

\begin{deluxetable}{rccccccc}
\tablecaption{Basic data of the identified knots}
\tablehead{
\colhead{No.} &
\colhead{Name} &
\colhead{$d$} &
\colhead{$D$} &
\colhead{$s$} &
\colhead{$S$} &
\colhead{$w$} &
\colhead{$W$} \\
\colhead{} &
\colhead{} &
\colhead{($^{\prime\prime}$)} &
\colhead{(kpc)} &
\colhead{($^{\prime\prime}$)} &
\colhead{(kpc)} &
\colhead{($^{\prime\prime}$)} &
\colhead{(kpc)} 
}
\startdata
1 & A1 & 5.8   & 3.6   & \nodata & \nodata & \nodata & \nodata  \nl
2 & A2 & 13.5  & 8.4   & 7.7  & 4.8  & \nodata & \nodata  \nl
3 & A3 & 24.8  & 15.5  & 11.3 & 7.1  & \nodata & \nodata  \nl
4 & A4 & 32.8  & 20.5  & 8.0  & 5.0  & \nodata & \nodata  \nl
5 & A5 & 51.7  & 32.4  & 18.9 & 11.9 & \nodata & \nodata  \nl
\hline
6 & B1 & 72.6  & 45.5  & \nodata & \nodata & 66.8  & 41.8  \nl
7 & B2 & 85.8  & 53.7  & 13.2 & 8.3  & 72.3  & 45.3  \nl
8 & B3 & 101.6 & 63.6  & 15.8 & 9.9  & 76.8  & 48.1  \nl
9 & B4 & 111.3 & 69.7  & 9.7  & 6.1  & 78.6  & 49.2  \nl
10 & B5 & 140.2 & 87.8  & 28.9 & 18.1 & 88.5  & 55.4  \nl
\hline
11 & C1 & 192.7 & 120.7 & \nodata & \nodata & 120.1 & 75.2 \nl
12 & C2 & 208.8 & 130.8 & 16.2 & 10.1 & 123.1 & 77.1  \nl
13 & C3 & 228.5 & 143.1 & 19.6 & 12.3 & 126.9 & 79.5  \nl 
14 & C4 & 243.5 & 152.5 & 15.0 & 9.4  & 132.1 & 82.8  \nl
15 & C5 & 271.2 & 169.9 & 27.7 & 17.3 & 131.0 & 82.1  \nl
\enddata
\end{deluxetable}

\newpage

\figcaption{
The top panel shows 
The radio continuum image at 4885 MHz obtained by Perley et al. (1984).
Three Chains A, B, and C together with the identification of radio knots
are separately shown in the lower three panels.
\label{fig1}}

\figcaption{
The definitions of the quantities used in this paper (see text).
\label{fig2}}

\figcaption{
The angular distances ($d$) of the 15 knots are plotted against 
the sequential knot number.
\label{fig3}}

\figcaption{
The separations of the $i$-th knots between two adjacent Chains
($w$).
\label{fig4}}

\figcaption{
The angular separations ($s$) between two adjacent knots are plotted 
against the sequential knot number.
\label{fig5}}

\figcaption{
The angular separations among the knots in  three Chains ($s$).
are shown as a function of the knot number.
\label{fig6}}

\figcaption{
The top two panels show 
the radio continuum image at 1662 MHz obtained by Perley et al. (1984).
The upper panel shows Chain A and the lower one shows Chain B.
Close-up images of the eight radio knots are shown in the lower
panel. Each arrow shows resolved substructure of the knot. 
\label{fig7}}

\figcaption{
A schematic illustration of the bended jet model for the 
radio jet of NGC 6251.
\label{fig8}}


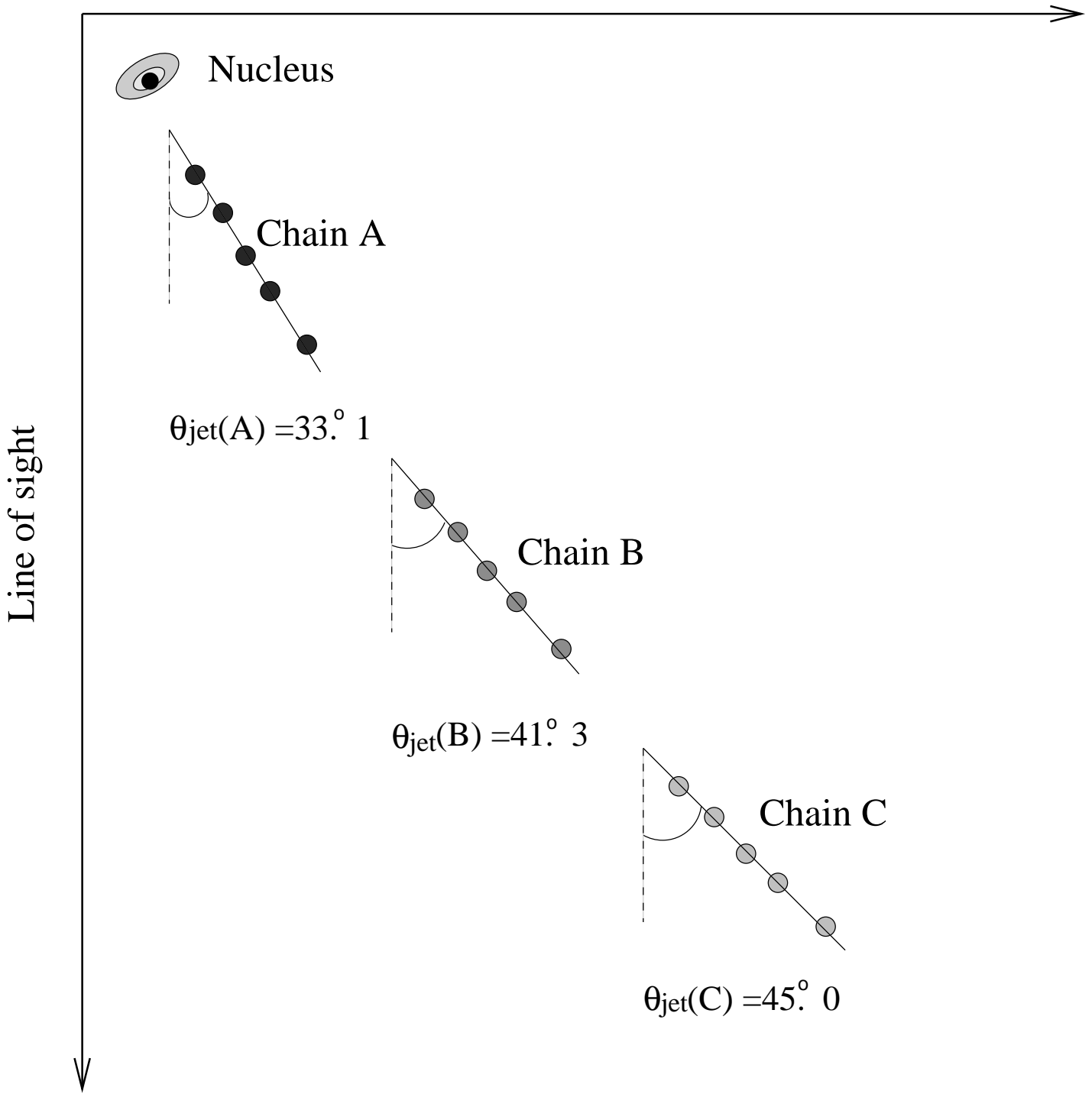
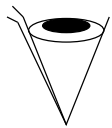
\end{document}